\documentclass[twocolumn,showpacs,prl,10pt]{revtex4}
\usepackage[english]{babel}
\usepackage{amsmath,amssymb}
\usepackage{times}
\usepackage{epsfig}

\begin{document}

\title {Double dispersion of the magnetic resonant mode in cuprates}
\author{I. Sega$^1$ and P. Prelov\v sek$^{1,2}$}
\affiliation{$^1$J.\ Stefan Institute, SI-1000 Ljubljana,
Slovenia}
\affiliation{$^2$Faculty of Mathematics and Physics, University
of Ljubljana, SI-1000 Ljubljana, Slovenia} 
\date{\today}

\begin{abstract}
The magnetic excitation spectra in the vicinity of the resonant peak,
as observed by inelastic neutron scattering in cuprates, are studied
within the memory-function approach. It is shown that at intermediate
doping the superconducting gap induces a double dispersion of the
peak, with an anisotropy rotated between the downward and upward
branch. Similar behavior, but with a spin-wave dispersion at higher
energies, is obtained for the low-doping case assuming a large pairing
pseudogap.

\end{abstract} 
\maketitle 

The magnetic resonant mode, first observed in the superconducting (SC)
phase of YBa$_2$Cu$_3$O$_{6+x}$ (YBCO) \cite{ross}, has been in the
last decade the subject of numerous studies, with the essential
information coming from the inelastic neutron scattering (INS)
experiments \cite{bour,fong}. It has been found that the peak
intensity is highest at the commensurate wavevector ${\bf
Q}=(\pi,\pi)$, while its frequency $\omega_r$ shifts with doping. More
recently, detailed studies of the magnetic response in the vicinity of
the resonant peak revealed several intriguing but quite universal
features.  While in the SC YBCO the stronger component of the resonant
mode disperses downwards \cite{bour1}, another branch apparently
emerging from the same peak shows upward dispersion
\cite{arai,rezn,pail}.  Similar features have been observed in the
underdoped YBCO, whereby the upper branch evolves into a
spin-wave-like mode at higher energies \cite{hayd,stoc}.  It is quite
remarkable that dispersions for various doping show quite consistent
anisotropic intensity within the ${\bf q}$ plane \cite{rezn,pail,hayd} with a 
rotation angle $45^0$ between the upper and lower branch.

On the theory side there appears to be a consensus that the resonant peak can
be interpreted as a low-energy collective antiferromagnetic (AFM) soft
mode, becoming undamped (at least underdamped) for $T<T_c$ due to the
onset of the $d_{x^2-y^2}$ SC gap in the electron-hole excitation
spectrum \cite{lava}. Two limits of the same scenario seem to be
realized. At optimum doping and slightly underdoped YBCO the resonant
mode is weak \cite{bour}, indicating that the collective mode is
weakly bound excitonic state within the SC gap
\cite{lava,morr,brin,norm,erem}. On the other hand, at low doping the
dominant part of the intensity of spin fluctuations with ${\bf q}={\bf
Q}$ are within the resonant peak, so the latter one is closer to an
undamped AFM paramagnon mode \cite{sega}.

The downward dispersion of the resonant mode is within the
random-phase approximation (RPA) and related theories for the
dynamical spin susceptibility $\chi_{\bf q}(\omega)$
\cite{lava,norm,chub,erem} a natural consequence of the closing of the
$d_{x^2-y^2}$ SC gap towards the nodal direction of the Fermi surface
(FS). The RPA seems to capture also some upward component (silent
band) after the disappearance of the downward branch \cite{erem}.

In this work we present results of the memory function approach to
spin dynamics \cite{sega}, which is capable to capture the upward
dispersion of the resonant mode. Moreover, it is applicable also in
the low doping regime. At the same time, memory function
representation offers an appropriate framework (broader than RPA) for
the general discussion of the INS experiments. Thus, it will be shown
that the explanation of collective mode properties at low doping
implies the existence of a large SC-like pseudogap. 

The dynamical spin susceptibility can be generally expressed in the
form \cite{sega}
\begin{equation}
\chi_{\bf q}(\omega)=\frac{-\eta_{\bf q}}{\omega^2+\omega M_{\bf
q}(\omega) - \omega^2_{\bf q}} \,, \label{chiq}
\end{equation}
where the 'spin stiffness' $\eta_{\bf q}=-{\dot\iota}\langle
[S^z_{-\bf q}\,\dot{S}^z_{\bf q})]\rangle $ can be evaluated for
within models relevant to cuprates \cite{sega}, while the 'mode frequency'
$\omega_{\bf q}=(\eta_{\bf q}/\chi^0_{\bf q})^{1/2}$ is related to the
static susceptibility $\chi^0_{\bf q}=\chi_{\bf q}(\omega=0)$. The
latter is a sensitive quantity, so we fix it with the
fluctuation-dissipation relation \cite{sega,prel}
\begin{equation}
\frac{1}{\pi}\int_0^\infty d\omega ~{\rm cth}\frac{\omega}{2T}
\chi^{\prime\prime}_{\bf q}(\omega)= \langle S^z_{-{\bf q}} S^z_{\bf
q}\rangle = C_{\bf q}\, , \label{eqsum}
\end{equation}
whereby the correlation function $C_{\bf q}$ is better known within
relevant models and also a more restricted quantity, although not
directly measured via INS so far. Depending on the damping function
$\Gamma_{\bf q}(\omega)= M^{\prime\prime}_{\bf q}(\omega)$,
Eq.~(\ref{chiq}) is able to deal with the overdamped response in the
normal state, with the spin-wave dispersion at higher energies (at low
doping) as well as with the resonant peak in the SC.

Using the method of equations of motion within the $t$-$J$ model it
has been shown that the collective spin fluctuations decay into
electron-hole excitations \cite{sega}. This leads to the lowest-order
mode-coupling approximation for the damping in the normal state
\begin{eqnarray}
\Gamma_{\bf q}(\omega) &=&\frac{\pi}{2\eta_{\bf q}\omega N} 
\int d\omega^\prime [f(\omega^\prime)-f(\omega+\omega^\prime)] 
\times \nonumber \\ 
&&\sum_{\bf k} w^2_{\bf kq}
A_{\bf k}(\omega^\prime) A_{{\bf k}+{\bf q}}(\omega+\omega^\prime) \, ,
\label{gamma} 
\end{eqnarray} 
where $w_{\bf kq}$ is the effective spin-fermion coupling \cite{sega}
and $A_{\bf k}(\omega)=$ is the single-particle spectral
function. Provided the existence of `hot spots' where the FS crosses the AFM
zone boundary (being the case for cuprates at low to intermediate
doping) we assume that at low-$\omega$ quasiparticles with dispersion
$\epsilon_{\bf k}$ and weight $Z_{\bf k}$ can determine the spectral function
$A_{\bf k}(\omega)=Z_{\bf k} \delta(\omega-\epsilon_{\bf k})$. This results
in a rather constant $\Gamma_{\bf q}(\omega)$ within the normal state at
low-$\omega$ and at ${\bf q} \sim {\bf Q}$. Although it is derived within
the specific prototype model, the form of Eq.~(\ref{gamma}) is quite generic
for the damping of the collective magnetic mode in a metallic system, since
the lowest-energy decay processes naturally involve the electron-hole
excitations close to the FS. It should be noted that similar expressions
appear also in theories based on the RPA approach \cite{morr,erem}.  For the
SC phase at $T<T_c$ Eq.~(\ref{gamma}) has to be generalized to include the 
anomalous spectral functions \cite{morr} leading to
\begin{eqnarray}
\Gamma_{\bf q}(\omega)&&\sim \frac{\pi}{2 \omega N} \sum_{\bf k}\tilde
w^2_{\bf kq} (u_{\bf k}v_{{\bf k}+{\bf q}}-v_{\bf k}u_{{\bf k}+{\bf
q}})^2\times\nonumber\\ &&[f(E_{\bf k})-f(E_{\bf k}-\omega)]\,
{\rm\delta}(\omega-E_{\bf k}-E_{{\bf k}+{\bf q}})\bigr] ,
\label{gamsc}
\end{eqnarray}
where $\tilde w^2_{\bf kq}=w^2_{\bf kq}Z_{\bf k}Z_{{\bf k}+{\bf
q}}/\eta_{\bf q}$, while $u_{\bf k},v_{\bf k}$ are the usual BCS
coherence amplitudes and $E_{\bf k}=\sqrt{\epsilon_{\bf
k}^2+\Delta^2_{\bf k}} $.

We are interested in the behavior for ${\bf q}\sim {\bf Q}$. So we use
for simplicity constant $\tilde w_{\bf kq} \sim {\bar w}$. Clearly, the most
relevant region is the vicinity of the 'hot spots' ${\bf k}_0$ where
FS crosses the AFM zone boundary. The latter depends on the effective
quasiparticle band, which we take of the form $\epsilon_{\bf k} = - 4 \eta_1 t
\gamma_{\bf k}- 4 \eta_2 t' \gamma'_{\bf k} -\mu$ where $\gamma_{\bf
k}=(\cos k_x+ \cos k_y)/2$, $\gamma'_{\bf k}=\cos k_x \cos k_y$ . In
the following, we assume values $\eta_1=\eta_2=0.33$, $t'/t=-0.33$
and $t \sim 400$~meV, corresponding roughly to hole-doped cuprates, in
particular YBCO at intermediate doping. The chemical potential $\mu$
is chosen so that the volume inside the Fermi surface corresponds to
Luttinger theorem at particular hole concentration $c_h$, i.e., $V_F
\propto 1-c_h$. $\eta_{\bf q}$ in Eq.~(\ref{chiq}) is well known from
model calculations \cite{sega} and quite restricted in range, and we
take $\eta_{\bf q} = 0.5~t$. For the SC gap we assume the
$d_{x^2-y^2}$ form, $\Delta_{\bf q}=\Delta_0( \cos q_x -\cos
q_y)/2$. Thus we end up with few adjustable parameters at chosen 
$c_h$: the correlation function $C_{\bf q}$, the effective coupling
$\bar w$ and the maximum SC gap $\Delta_0$.

{\it Intermediate - optimum doping}: Within this regime the collective
mode is heavily overdamped in the normal state. The indication for the
latter is low  intensity of the INS in the
relevant low-energy window. Following the sum rule, Eq.~(\ref{eqsum}),
this can be made compatible only with modest $C_{\bf q}\lesssim 1$. Furtheron
we assume the Lorentzian form $C_{\bf q}=C_{\bf
Q}/(1+\tilde q^2 /\kappa^2)$ where $\tilde {\bf q}={\bf q}-{\bf Q}$. To
be specific, we fix for the presented case the 'optimum' doping at
$c_h=0.15$ with $C_{\bf Q}= 1.0$ and $\kappa\sim 1.25$. The SC gap is
roughly known from experiments and  we take $\Delta_0=40$~meV. The
remaining  input is the coupling $\tilde w$ or equivalently $\Gamma_{\bf Q}$
within the normal state.  We have shown in our analytical derivation within
the $t$-$J$ model \cite{sega} that $\Gamma_{\bf q} \propto t$, but it is
essentially renormalized due to AFM spin correlations, in particular at low
doping. We can stress that for the appearance of the upper resonant branch
it is crucial that $\Gamma_{\bf Q}$ is not too large, as seems to be
inherent within the RPA \cite{morr,erem} which otherwise yields within the
intermediate-doping regime formally quite similar expressions to our
Eqs.~(\ref{chiq}),(\ref{gamsc}). For results at intermediate doping we use
below $\Gamma_{\bf Q}\sim 0.45t$. 

\begin{figure}[htb]
\centering \center{\epsfig{file=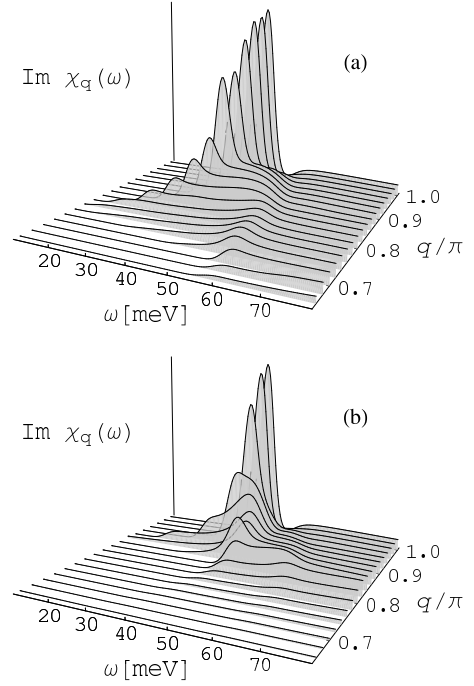,width=61mm}}\\
\caption{Magnetic fluctuations spectra $\chi^{\prime\prime}_{\bf q}(\omega)$
(arbitrary units) at intermediate doping $c_h=0.15$ for momenta: a) along the $x$
direction ${\bf q}=q(1,0)$, and b) along the zone diagonal ${\bf
q}=q(1,1)$.}
\label{fig1}
\end{figure}

The spectra in the vicinity of the resonance $\chi^{\prime\prime}_{\bf
q}(\omega \sim \omega_r)$ for ${\bf q}\sim {\bf Q}$ have been partly
studied in Ref.~\cite{sega}. Presented results show besides a
pronounced downward dispersion also a weaker upward
branch. In Fig.~1 we display $\chi^{\prime\prime}_{\bf q}(\omega)$ for
momenta  both along the $x$-axis, ${\bf q}=q(1,0)$, and along the zone
diagonal  ${\bf q}=q(1,1)$, while in Fig.~2 we present the planar ${\bf q}$
scans of the intensity $\chi^{\prime\prime}_{\bf q}(\omega)$ at fixed
$\omega$, as frequently employed in presentation of INS results.

\begin{figure}[htb]
\centering
\center{\epsfig{file=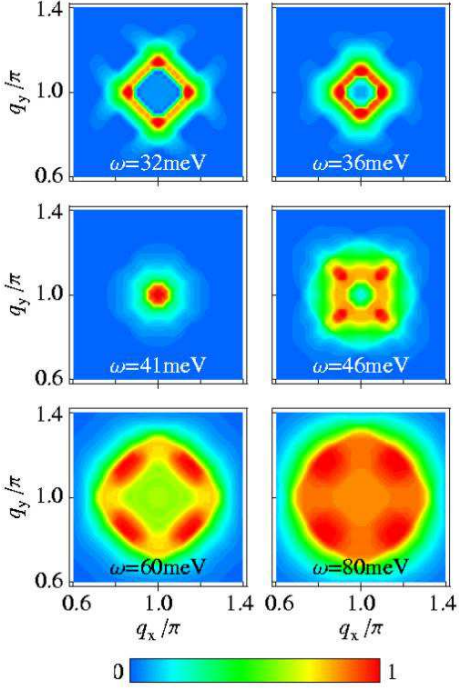,width=61mm}}\\
\caption{Normalized intensity plot of $\chi^{\prime\prime}_{\bf q}(\omega)$
in the ${\bf q}$ plane at intermediate doping for selected energies
$\omega$ below and above the resonant peak at $\omega_r\sim 41$~meV.}
\label{fig2}
\end{figure}
 
Following observations can be made on the basis of Figs.~1,2: a) both
presentations clearly reveal two branches emerging from the same
coherent resonant mode at $\omega_r \sim 41$~meV. Intensity plots of both
branches within the ${\bf q}$ plane are square-like around AFM ${\bf
Q}$ (see Fig.2), however with quite pronounced anisotropy.  b) For the
downward  branch the intensities are strongest along the $(1,0)$ direction.
This is consistent with the faster dispersion along the zone diagonal
$(1,1)$ (see Fig.~1a)  which reduces intensity relative to the $(1,0)$
direction and deforms the constant $\omega$-scan into a square-like
pattern. c) The development is more sensitive for $\omega>\omega_r$,
still the situation with the upward branch is just opposite to the
downward one. The dispersion is stronger along the $(1,0)$ direction
and consequently the larger intensity is  along the $(1,1)$ direction. d)
Above  the damping threshold $\omega>~2\Delta_0$ the upward branch merges
into  an incoherent response broad both in ${\bf q}$ as well as in $\omega$.
Note,  however, that the incoherent part still exhausts most of the
intensity sum  rule, Eq.~(\ref{eqsum}), even for ${\bf q}={\bf Q}$.
 
Let us give some explanation for the behavior of the collective mode
as observed in Figs.~1,2. At intermediate (near optimum) doping the
normal-state damping is large, $\Gamma_{\bf Q}>\omega_{\bf Q}$, and
the collective mode is heavily overdamped in the normal state. The
sharp resonant peak at ${\bf Q}$ appears due to the step-like
vanishing damping $\Gamma_{\bf Q}(\omega<\Omega_{\bf Q})=0$ within the
SC phase, where $\Omega_{\bf Q}=2\Delta_{{\bf k}^*}$ and ${\bf k}^*$
represents the location of the 'hot spot' on the FS. Since the damping
cut-off is below the characteristic 'mode frequency', $\Omega_{\bf
Q}<\omega_{\bf Q}$, the character of the resonant mode is
excitonic-like \cite{lava,morr}. I.e., it appears lower but close to
$\Omega_{\bf Q}$ and consequently carries only a small part of the
whole sum rule, Eq.~(\ref{eqsum}) \cite{sega}.  The dispersion of the
mode, both the downward \cite{chub,erem} as well as the upward one, is
intimately related to the properties of the SC gap $\Delta_{\bf
k}$. As noted before \cite{morr,sega,erem}, the damping function
$\Gamma_{\bf q}(\omega)$ shows for ${\bf q}\ne {\bf Q}$ several steps,
in contrast to a single step at ${\bf Q}$. Thresholds are determined by
the 'hot spot' condition, i.e., by processes of zero-energy
electron-hole excitations (in the normal state) connecting Fermi
surfaces ${\bf k}_{F1}+{\bf k}_{F2}={\bf q} +{\bf K}$ where ${\bf
k}_{Fi}$ are wavevectors on the FS and ${\bf K}$ are reciprocal
lattice vectors. Within the SC phase this leads to steps in the damping at
$\Omega^i_{\bf q}=|\Delta_{{\bf k}_{F1}}|+ |\Delta_{{\bf
k}_{F2}}|$. Away from ${\bf q}={\bf Q}$ there are in general four
nontrivial $\Omega^i_{\bf q}, i=1,4$ with a possible degeneracy for
${\bf q}$ with a higher symmetry in the Brillouin zone.

The lowest step at $\Omega^1_{\bf q}$ pushes the downward resonant
branch as $\omega_r({\bf q})<\Omega^1_{\bf q}$. The latter should for
${\bf q}=q(1,1)$ approach zero at ${\bf q}_n=2{\bf k}_{Fn}$ where
${\bf k}_{Fn}$ is the nodal point on the FS. It is, however, clear
from Fig.~1b that the branch loses intensity before reaching this
${\bf q}_n$. The dispersion of $\Omega^1_{\bf q}$ along the $(1,0)$
direction is substantially weaker, as seen in Fig.~1a, which leads to
square-like contours and the $(1,0)$ dominated anisotropy in Fig.~2.
The upper branch in our analysis appears as a exciton-like resonance
below next thresholds, i.e., $\omega_u({\bf q})<\Omega^i_{\bf q},
i>1$. The condition for such a resonance is that the finite
damping $\Gamma_{\bf q}(\omega_u)$ is not too large. The latter seems
to be the case within the RPA analysis \cite{erem}, where the upper
branch does not emerge from the resonant peak. Finally, for
$\omega>2\Delta_0$ the damping $\Gamma_{\bf q}(\omega)$ is large and
quite constant and resonant features disappear, with the remaining
strongly overdamped AFM paramagnon mode.

\begin{figure}[!htb]
\centering
\epsfig{file=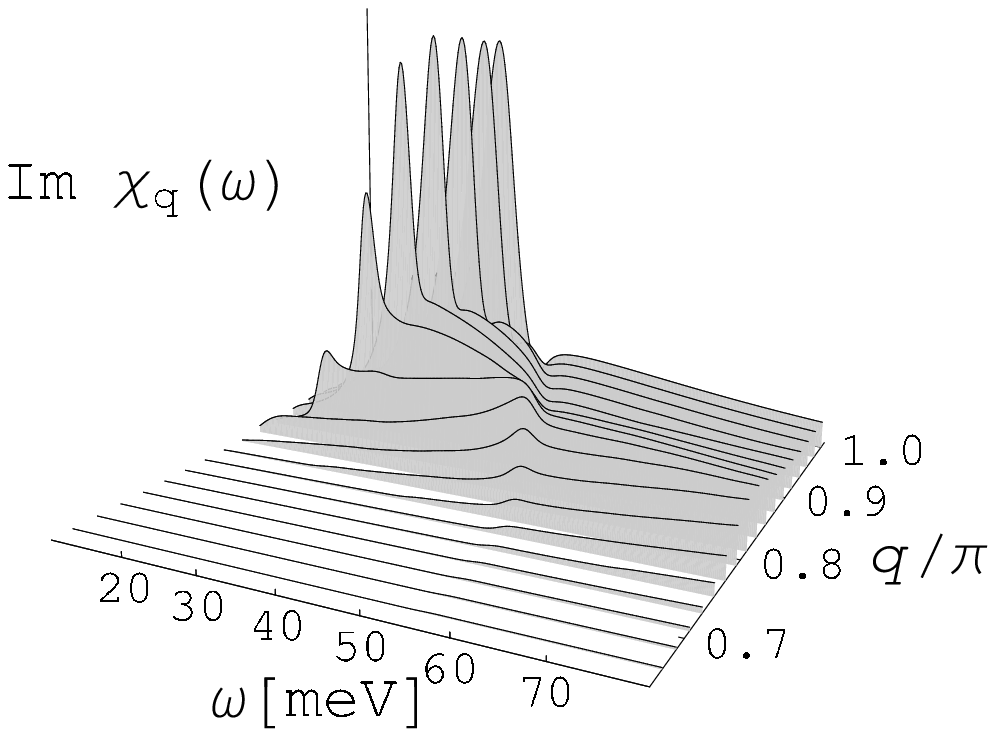,width=61mm}
\caption{$\chi^{\prime\prime}_{\bf q}(\omega)$ (arbitrary units) at low
doping $c_h=0.1$ for ${\bf q}=q(1,0)$.}
\label{fig3}
\end{figure}

{\it Low-doping}: At low doping one expects smaller normal-state
damping $\Gamma_{\bf Q}(\omega)$ \cite{sega} but at the same time
larger $C_{\bf Q}$, which generates spin-wave-like dispersion at
larger $\omega$, as observed in INS \cite{hayd,stoc}. Both facts also
lead to lowering of $\omega_{\bf Q} \propto c_h$ \cite{sega}, which at
the same time corresponds closer to the resonance $\omega_r \sim
\omega_{\bf Q}$. Moreover, the resonance peak exhausts substantial part of the 
sum rule.  To account for a weak anisotropy as well as the
spin-wave-like dispersion, we use in the calculation for $C_{\bf q}$
the form which incorporates an incommensurability $\delta_i=(\pm \delta, 0)$
and $(0,\pm \delta)$ 
\begin{equation}
C_{\bf q}= A \sum_i^4 \frac{1}{\sqrt{\kappa^2+3|{\bf q}-{\bf Q}_i|^2}},
\label{cq}
\end{equation}
where ${\bf Q}_i={\bf Q}+\delta_i$.

In contrast to the intermediate doping, a direct application of the
damping, Eq.~({\ref{gamma}), with a single SC gap seems not to be
sufficient to describe the observed INS results. First, the resonant peak in
underdoped YBCO remains broad (not resolution limited) for $T<T_c$,
only compatible with a finite damping persisting in the SC phase
\cite{bour,fong,stoc}. Still, there is some signature of a 
double dispersion \cite{stoc}, although the downward dispersing mode
is much less pronounced. Also, INS results reveal a drop of intensity
for $\omega<\omega_c$, where $\omega_c<\omega_r$ can be interpreted as
the spin-gap energy scale \cite{bour,fong} showing up also in NMR
relaxation experiments.

 
\begin{figure}[htb]
\centering
\epsfig{file=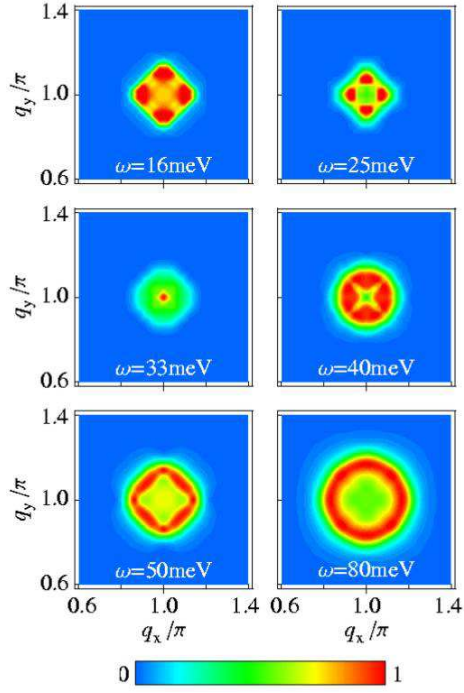,width=61mm}
\caption{Normalized intensity plot of $\chi^{\prime\prime}_{\bf q}(\omega)$
in the $q$ plane at low doping. The resonant peak is at $\omega_r\sim 33$~meV.}
\label{fig4}
\end{figure}

To account for these observations, we generalize at low doping the damping
function $\Gamma_{\bf q}(\omega)$, Eq.~(\ref{gamsc}), as
follows. The SC gap $\Delta_{\bf q}$ is replaced by a large pseudogap
$\Delta^*_{\bf q}$ with the same $d_{x^2-y^2}$-character and resulting
$\Gamma^*_{\bf q}(\omega)$. It is well possible that the latter behavior
persists also above $T>T_c$, but below the pseudogap temperature $T<T^*$,
well established for underdoped cuprates.  However, there is an
additional damping, $\Gamma_{\bf q}(\omega)=\Gamma^*_{\bf
q}(\omega)+\Gamma^c_{\bf q}(\omega)$, whereby $\Gamma^c_{\bf
q}(\omega<\Omega^i_{\bf q})$ remains finite even within the (pseudo)
SC gap. Still, there exists a lower scale, possibly the coherent SC (or
spin) gap, below which $\Gamma^c_{\bf q}(\omega<\omega_c)=0$. Since
$\omega_c<\omega_r$ the dispersion of $\Gamma^c_{\bf q}(\omega>\omega_c)$
does not have any significant effect and we assume further for simplicity a
constant $\Gamma^c$.
    
In Figs.~3,4 we display results for low $T \sim 0$, corresponding to
low doping $c_h \sim 0.1$, where following parameters have been
adopted: $C_{\bf Q}= 1.6$, $\kappa = \delta \sim 0.3$, damping
$\Gamma^c =18$~meV, $\Gamma^*_{\bf Q}=60$~meV, and gaps
$\Delta_0^*=38.5$~meV, $\omega_c=10$~meV. While there are similarities with
the  intermediate doping results in Figs.~1,2, there are also evident
differences. The resonant peak is broader, but underdamped, due to finite
$\Gamma^c<\omega_r$.  There is a signature of a downward branch with
the same anisotropy as for the intermediate doping, but this branch is
much less pronounced and losing fast in intensity. The upward
dispersion is stronger and transforms at larger $\omega$  into
the usual isotropic AFM spin-wave dispersion with $\omega \propto
\tilde q$. 

Let us discuss the correspondence of presented results with INS
experiments on cuprates. For the intermediate doping our results agree
with features seen in optimum or slightly underdoped YBCO
\cite{arai,pail,rezn}: a) pronounced downward dispersion with an enhanced 
intensity along the $(0,1)$ direction, b) broader and less pronounced
upward dispersive branch with stronger intensity along the $(1,1)$
direction. Both branches with quite similar anisotropy we obtain also
for low doping  in Figs.~3,4, whereby the upper one evolves into
spin waves at higher $\omega$. This seems to be in agreement
with  underdoped YBCO \cite{hayd,stoc}. However, it should be
pointed out that INS reveals in the latter case a  much less pronounced
downward branch, in contrast to optimum doping.

In conclusion, we have shown that the resonant peak double dispersion
and its anisotropy are a notrivial consequence of the damping
$\Gamma_{\bf q}(\omega)$ in the SC phase, reflecting the $d$-wave gap
$\Delta_{\bf q}$ structure and related thresholds $\Omega^i_{\bf
q}$. In this sense, the INS results on resonant peak serve as a very
stringent test for the mechanism of the collective mode decay and
the structure of the SC gap. While our basic assumption \cite{sega} of the
decay into electron-hole excitations, consistent with other authors
\cite{morr, erem}, does not offer much freedom of interpretation at 
intermediate doping, there are several open issues at low doping.  We
get a reasonable explanation of experiments only after assuming two
distinct energy scales, i.e., the low spin-gap $\omega_c$ and the
large $d$-wave-like pseudogap $\Delta_{\bf q}^*$. Without the latter
it would be quite hard to account for two branches emerging from
a single resonant feature at $\omega_r$ and commensurate ${\bf Q}$. We should
point out that a similar quasi-universal development is observed even
in non-SC cuprates \cite{tran}. Thus, experiments on magnetic collective
mode dispersion, in particular at low doping, can get a new insight on
the long-standing question of low-energy excitations in pseudogap phase of
cuprates.

\end{document}